# Thinned Germanium Substrates for III-V Multijunction Solar Cells

Ivan Lombardero[1], Naoya Miyashita[2], Mario Ochoa[1,3], Yoshitaka Okada[2], Carlos Algora[1]

[1]Instituto de Energía Solar, Universidad Politecnica de Madrid, 28040 Madrid, Spain
[2]Research Center for Advanced Science and Technology (RCAST), The University of Tokyo, 4-6-1 Komaba, Meguro-ku, Tokyo 153-8904, Japan
[3]Laboratory for Thin Films and Photovoltaics, Empa-Swiss Federal Laboratories for Materials Science and Technology, Ueberlandstrasse 129, 8600 Duebendorf, Switzerland

*Abstract*—Multijunction solar cells are usually grown on Ge substrates. This implies several disadvantages that hinder the performance of the whole multijunction and limit their possible applications. The drawbacks caused by the substrate are: heavier devices, higher operation temperatures, lower performance and lack of photon confinement. In this work we propose thinning the substrate as a valid solution to the aforementioned challenges. The influence of the substrate thickness on the Ge subcell performance inside a multijunction is simulated using 2D TCAD tools. Simulation results point to the back surface recombination as the key parameter to enhance the development of thinned Ge subcells. Ge substrates have been thinned down, achieving 115μm thick samples. Finally, solar cells have been manufactured out of the thinned substrates proving a limited degradation and showing the feasibility of this process to manufacture Ge subcells thinned down up to 115μm.

*Index Terms*—thin solar cells, chemical thinning, III-V solar cells, space solar cells, germanium.

## I. Introduction

Semiconductor substrates are used as the solar cell base in certain structures, among we can find germanium solar cells. Mostly used as multijunction's bottom subcell, Ge solar cells are usually fabricated on p-doped substrates by phosphorous diffusion inside a metal-organic vapor phase epitaxy (MOVPE) reactor. Although most of the efforts trying to improve multijunction solar cells performance focused on increasing the number of junctions [1]–[3], there are several drawbacks caused by the substrate that limits their performance. Among them we can find heavier devices, higher operation temperatures, lower voltage generation and lack of photon confinement. The use of thinned substrates is proposed to solve, or at least mitigate, these detrimental effects. This would allow to improve the current technology without involving any major changes on the semiconductor structure nor the device design either. Therefore, the aim of this study is to point out the advantages of a thinned Ge subcell for multijunction solar cells with special emphasis on triple junction ones.

## II. Substrate Influence Assessment

Ge substrates for multijunction solar cells are usually 165-185μm thick and p-type doped $10^{17}$-$10^{18}$ cm$^{-3}$. The optimum doping level is derived from the Ge minority carrier properties (which strongly depends on the doping level[4], [5]) and its resistivity. The thickness is determined by the manufacturing process, which requires a minimum thickness to ensure a high yield. Typical substrates doped 7·$10^{17}$ cm$^{-3}$ with a thickness of 175 μm will be assumed in this work.

### A. Solar Cell Weight

The solar cell weight is of paramount importance for some applications where the weight dramatically influences the cost, such as in space applications[6]. More than 95% of the total weight of a standard GaInP/Ga(In)As/Ge triple junction is caused by the Ge substrate, which highlights the importance of making it thinner. Fig.1 shows the weight of a triple junction solar cell as a function of the substrate thickness assuming 2μm and 5μm for the GaInP and Ga(In)As subcells respectively. These thicknesses are overestimated to take into account other layers apart from the subcells themselves, such as the tunnel junctions or the buffer layers. It can be seen that once the substrate is thinned down to 10μm the weight reduction starts to saturate, achieving values lower than 10% of the total weight.

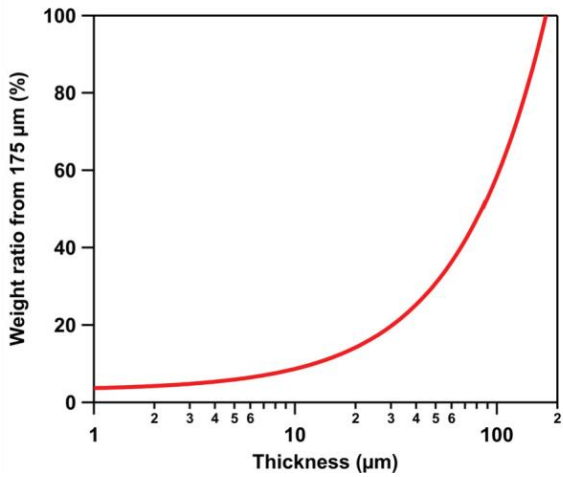

Fig. 1. GaInP/Ga(In)As/Ge triple junction solar cell weight reduction as a function of the Ge substrate thickness. 2μm and 5μm thick layers for the GaInP and Ga(In)As are consider respectively to take into account other layers apart from the subcells themselves. The initial substrate thickness considered is 175μm.

### B. Heat Absorption

Ge substrates suffers from free carrier absorption (FCA)[7], absorbing wavelengths longer than their bandgap and heating up the solar cell[8]. This kind of absorption becomes noticeable once the photogeneration is dominated by indirect transitions (λ>1600nm). Consequently, most of the light beyond 1600nm will be transformed into heat, degrading the device performance. Fig.2 shows the spectral irradiance for AM0, AM1.5G and AM1.5D spectra, together with their cumulative irradiance (i.e. the irradiance integral from 300nm).

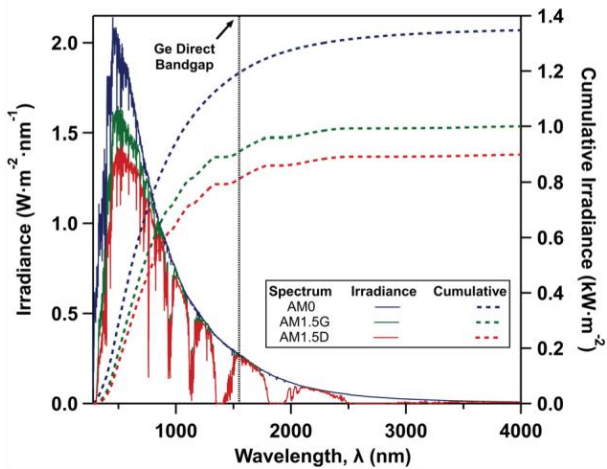

Fig. 2. AM0, AM1.5G and AM1.5D spectral irradiance. The cumulative irradiance (i.e. integral from 300nm) for each spectrum is also shown.

The irradiance beyond 1600nm is around 8% of the total energy for each spectrum. If we consider that the other 90% is transformed into electricity with an efficiency around 40% for the standard triple junction, wavelengths longer than 1600nm accounts for 14% of the total irradiance heating up the solar cell.

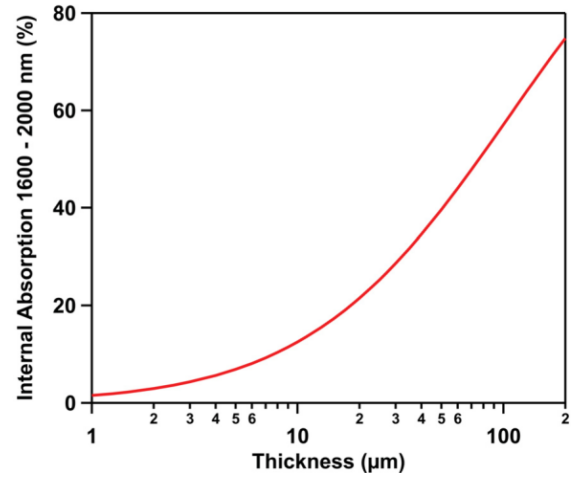

Fig. 3. Internal absorption (see equation 1) for a triple junction as a function of the substrate thickness.

In order to assess how thin the substrate should be to avoid being heated up by useless wavelengths, the absorption between 1600 and 2000nm has been simulated for different substrate thicknesses (see Fig.3). The transfer matrix method (TMM) was used to perform the optical calculations using data from [9]. To avoid the influence of optical effects, the internal absorption has been calculated:

$$A_{Internal}(\lambda) = A(\lambda)/(1 - R(\lambda)) \qquad (1)$$

where $A$ stands for the Absorptivity and $R$ for the Reflectivity. Fig.3 shows that the absorption decrease starts to saturate for thicknesses around 10μm. For this thickness the absorption has decreased from 71 to less than 13%, which means that the heat absorption for long wavelengths would decrease by more than 80%.

### C. Performance at 1 Sun

Now, we proceed to evaluate the performance of a Ge subcell. To do so, 2D simulations have been carried out with Silvaco Atlas[10], [11]. Neither shadowing nor resistive losses at the contacts have been considered. Optical calculations have been carried out using the Transfer Matrix Method (TMM) assuming no antireflection coating (ARC) and a gold back metal contact. The change in the operation temperature as a result of a lower heat absorption, caused by a thinner substrate, has not been considered.

An n/p Ge single junction with $10^{19}$ and $6 \cdot 10^{17} cm^{-3}$ constant doping levels for the emitter and the base respectively has been simulated. The emitter is assumed to be 190nm thick, being the rest of the substrate the base of the solar cell. A standard surface recombination velocity of $2\cdot10^5$ cm/s [12] has been considered at the window-emitter interface. To model the back surface recombination the infinite recombination at the back metal contact has been avoided by means of an ideal heterojunction at the back surface with a 1eV bandgap material. The affinity has been set to be such that there is no barrier for holes (i.e. at the valence band) forming a 375meV barrier at the conduction band. Then, the desired surface recombination velocity for electrons and holes at that interface has been set, considering $10^9$ cm/s as infinite recombination.

Fig.4 shows the evolution of the power generation of a Ge subcell under AM0 filtered by a GaAs layer for three different scenarios: only standard front surface recombination (*Front*), only infinite back surface recombination (*Back*) and combination of standard front and infinite back surface recombination (*Front and Back*). Thick substrates are dominated by the front surface recombination while thin ones are dominated by the back one. This different behavior is explained as the emitter thickness is constant and it does not change as the substrate is thinned down. Conversely, the thinner the substrate the closer the back surface is to the *pn* junction increasing its influence on the overall performance. For the optimum thickness pointed out in the previous sections (∼10μm), it is more important to avoid the back surface recombination than the front one in order to achieve a good performance in the Ge subcell. Moreover, as long as the back surface recombination is avoided, the power loss is limited to only 6% if the substrate is thinned down from 175 to 10μm.

### D. Photon confinement

Typical Ge substrates are not suitable to exploit the benefits of confining photons using a back reflector[13], [14]. First, the FCA will absorb most of the long wavelength photons, which are the ones that would benefit from the back reflector. Second, wavelengths influenced by the back reflector would be mostly absorbed too far away from the *pn* junction to be collected, given the typical diffusion length of electrons in highly doped germanium substrates (<100μm). Therefore, the collection of carriers photogenerated by long wavelengths are hindered by a thick, highly doped substrate. These two effects are avoided as the substrate is thinned down, improving the effectiveness of the back reflector.

## III. EXPERIMENTAL

Once the advantages of a thinner substrate for a multijunction solar cell has been evidenced, we proceed to assess how the thinning process could be performed and demonstrate its suitability to manufacture solar cells.

### A. Thinning Process

Ge substrates have been thinned down by means of chemical etching processes. This method has been selected due to its scalability and ease of application during the manufacture of a solar cell. Among the options reported to etch Ge substrates [15]–[19], acid-base etchants has been selected to carry out the etching process. In order to measure the etch rate one side of the sample was protected with photoresist. Then, the step was measured with the help of a profilometer. All etching processes were carried out at ambient temperature (23-27 °C) with enough solution in order to avoid its saturation. Standard (100) Ge substrates with a miscut of 6° towards the nearest (111), a thickness of 175μm and $6\cdot10^{17}$ cm$^{-3}$ doping level were used to assess the etch rate evolution with time as depicted in Fig.5. The etch rate is fairly constant around 10μm/hour, similar to what has been reported in the literature. Etching processes as long as 6 hours have been carried out achieving substrates thinned down to 115μm.

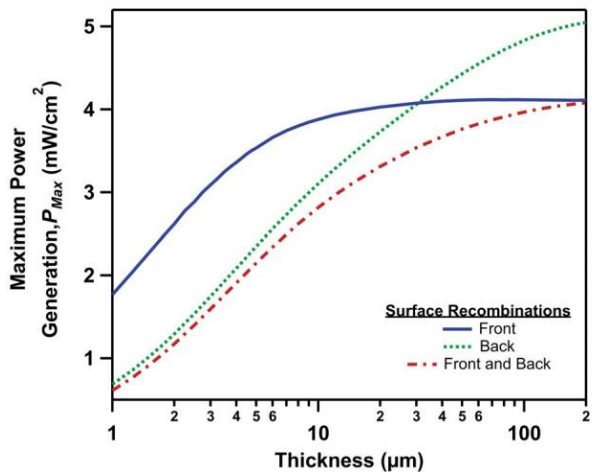

Fig. 4. Maximum power generated by a Ge solar cell under AM0 filtered by a GaAs layer. Three scenarios are plotted regarding the Ge solar cell surface recombinations: only standard window-emitter surface recombination (*Front*), only infinite back surface recombination (*Back*) and combination of standard window-emitter and infinite back surface recombination (*Front and Back*).

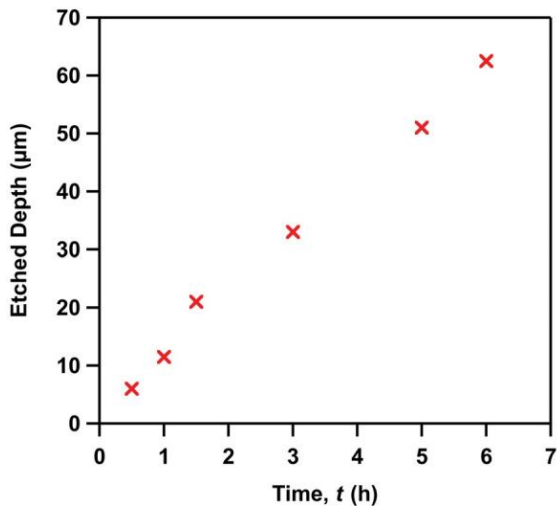

Fig. 5. Ecthed depth as a function of time for Ge substrates.

### B. Device Characterization

Germanium solar cells have been manufactured out of the thinned substrates. The illumination I-V curve of such devices has been measured under no strict spectrum control conditions. Table.I depicts average values for the main parameters ($V_{OC}$, $I_{SC}$, $FF$ and $P_{Max}$) of the manufactured solar cells for three different thicknesses: 175, 145 and 115$\mu$m. A clear trend can be observed for all parameters except for the $FF$. A degradation around 1.4% for $V_{OC}$ and 4.6% for $I_{SC}$ is detected when thinning the substrate from 175 to 115$\mu$m. $FF$ is almost unaffected by the sample thickness, and the small variations observed are mostly related to the number of fingers in the metal grid. Accordingly, the power generation decreases by a 6.3% as a result of the lower $V_{OC}$ and $I_{SC}$ in the 115 $\mu$m thick solar cell. Nonetheless, this result demonstrates that a solar cell can be thinned down to 115$\mu$m with limited losses, even in the presence of a high back surface recombination velocity.

TABLE I. Summary of I-V curve parameters

| Thickness | $V_{OC}$ | $I_{SC}$ | FF | $P_{Max}$ |
|---|---|---|---|---|
| ($\mu$m) | (V) | (mA) | (%) | (mW) |
| 175 | 0.2587 | 49.0 | 0.682 | 8.65 |
| 145 | 0.2572 | 47.7 | 0.679 | 8.33 |
| 115 | 0.2552 | 46.7 | 0.680 | 8.1 |

Simulations pointed to a degradation of 6% and 2% for the "*Back*" and "*Front and Back*" scenarios respectively (see Fig.4). This could point to an overestimated front degradation. However, this is only an hypothesis and a more thorough analysis regarding the difference between simulation and measurement results is required.

### IV. CONCLUSIONS

In this work, the drawbacks caused by the Ge substrate in the solar cell performance have been pointed out. Thinning the substrate has been demonstrated as a solution to decrease the weight of the solar cell, cool down the operating temperature, enhance the performance and allow to benefit from a back reflector. Thinning the substrate down to 10$\mu$m would reduce the weight by more than 90% while limiting the heat absorption in useless wavelengths (>1600nm) by more than 80%. In order to achieve a good performance for thinned Ge solar cells it is mandatory to avoid the back surface recombination while the front one will only degrade the performance by a 6% for 10$\mu$m thick samples. Ge substrates have been thinned down, achieving thicknesses around 115$\mu$m. Thinned Ge solar cells have been manufactured with limited losses, showing the feasibility of this process.

### ACKNOWLEDGMENTS

This work has been supported by the Fundacion Iber-´drola Espana Research Grants, Spanish MINECO through˜ the project TEC2017-83447-P and by the Comunidad de Madrid through the project MADRID-PV2 (S2018/EMT4308). I. Lombardero acknowledges the financial support from the Spanish Ministerio de Educacion, Cultura y Deporte´ through the Formacion del Profesorado Universitario grant´ with reference FPU14/05272.